\documentclass[aps,prb,showpacs,superscriptaddress,twocolumn]{revtex4-1}

\usepackage{amsmath,graphics}
\usepackage[next]{inputenc}
\usepackage[dvips]{epsfig}
\usepackage[colorlinks=true,citecolor=blue,linkcolor=blue]{hyperref}
\usepackage{bbm}
\usepackage{bbold}
\usepackage{booktabs}
\usepackage{multirow}
\usepackage{hhline}
\usepackage{graphicx}
\usepackage{amsmath}
\usepackage{multirow}
\usepackage[usenames, dvipsnames]{color}
\usepackage{color,soul}
\usepackage{latexsym}

\usepackage{subfigure}
\usepackage{hyperref}
\usepackage{float}
\usepackage{amssymb}
\usepackage{dcolumn}
\usepackage{bm}

\usepackage{graphicx}
\usepackage{subfigure}
\usepackage{amsmath}
\usepackage{xcolor}
\usepackage{color, soul}

\begin{document}


\title{Bulk Fermi surfaces of the Dirac type-II semimetallic candidate NiTe$_2$}

\author{Wenkai Zheng}
\email{wzheng@magnet.fsu.edu}
\affiliation{National High Magnetic Field Laboratory, Florida State University, Tallahassee, Florida 32310,USA}
\affiliation{Department of Physics, Florida State University, Tallahassee, Florida 32306,USA}

\author{Rico Sch\"{o}nemann}
\affiliation{National High Magnetic Field Laboratory, Florida State University, Tallahassee, Florida 32310,USA}
\altaffiliation{Present address: Los Alamos National Laboratory, Los Alamos, New Mexico 87545, USA}

\author{Shirin Mozaffari}
\affiliation{National High Magnetic Field Laboratory, Florida State University, Tallahassee, Florida 32310,USA}

\author{Yu-Che Chiu}
\affiliation{National High Magnetic Field Laboratory, Florida State University, Tallahassee, Florida 32310,USA}
\affiliation{Department of Physics, Florida State University, Tallahassee, Florida 32306,USA}

\author{Zachary Bryce Goraum}
\affiliation{National High Magnetic Field Laboratory, Florida State University, Tallahassee, Florida 32310,USA}
\affiliation{Department of Physics, Florida State University, Tallahassee, Florida 32306,USA}

\author{Niraj Aryal}
\affiliation{National High Magnetic Field Laboratory, Florida State University, Tallahassee, Florida 32310,USA}
\affiliation{Department of Physics, Florida State University, Tallahassee, Florida 32306,USA}
\email{Present address: Condensed Matter Physics and Materials Science Divison, Brookhaven National Lab, Upton, New York 11973,USA}

\author{Efstratios Manousakis}
\affiliation{National High Magnetic Field Laboratory, Florida State University, Tallahassee, Florida 32310,USA}
\affiliation{Department of Physics, Florida State University, Tallahassee, Florida 32306,USA}

\author{Theo M. Siegrist}
 \affiliation{National High Magnetic Field Laboratory, Florida State University, Tallahassee, Florida 32310,USA}
\affiliation{Department of Chemical and Biomedical Engineering, FAMU-FSU College of Engineering, Tallahassee, Florida 32310, USA}

\author{Kaya Wei}
 \affiliation{National High Magnetic Field Laboratory, Florida State University, Tallahassee, Florida 32310,USA}

\author{Luis Balicas}
  \email{balicas@magnet.fsu.edu}
\affiliation{National High Magnetic Field Laboratory, Florida State University, Tallahassee, Florida 32310,USA}
\affiliation{Department of Physics, Florida State University, Tallahassee, Florida 32306,USA}

\date{\today}

\begin{abstract}
Here, we present a study on the Fermi-surface of the Dirac type-II semi-metallic candidate NiTe$_2$ via the temperature and angular dependence of the de Haas-van Alphen (dHvA) effect measured in single-crystals grown through Te flux. In contrast to its isostructural compounds like PtSe$_2$, band structure calculations predict NiTe$_2$ to display a tilted Dirac node very close to its Fermi level that is located along the $\Gamma$ to A high symmetry direction within its first Brillouin zone (FBZ).
The angular dependence of the dHvA frequencies is found to be in agreement with the first-principle calculations when the electronic bands are slightly shifted with respect to the Fermi level ($\varepsilon_F$), and therefore provide support for the existence of a Dirac type-II node in NiTe$_2$. Nevertheless, we observed mild disagreements between experimental observations and density Functional theory calculations as, for example, nearly isotropic and light experimental effective masses. This indicates that the dispersion of the bands is not well captured by DFT. Despite the coexistence of Dirac-like fermions with topologically trivial carriers, samples of the highest quality display an anomalous and large, either linear or sub-linear magnetoresistivity. This suggests that Lorentz invariance breaking Dirac-like quasiparticles dominate the carrier transport in this compound.
\end{abstract}

\maketitle


\section{\label{sec:level1}INTRODUCTION}

In recent years, electronic topology has emerged as an important set of concepts shedding new light onto solid-state systems \cite{Review1,Review2,Review3}.
Predictions for novel electronic behavior include topologically protected surface states in topological insulators \cite{Hsieh2008,Qian1344}, quasiparticles whose behavior is akin to relativistic Dirac fermions in graphene \cite{Novoselov2005}, or chiral Weyl fermions, as uncovered in the family of TaAs compounds \cite{Xu613,PhysRevX.5.031013}, and even Majorana like fermions \cite{PhysRevLett.100.096407} that are predicted to be relevant for quantum computation \cite{Majoranas}.

Dirac type-I semimetallic systems are characterized by linearly dispersing electronic bands that touch at a point, the Dirac node. If these spin-degenerate bands crossed the Fermi level they would lead to near massless quasiparticles displaying high mobilities as in graphene. More recently, several transition metal dichalcogenide (TMD) compounds, including PdTe$_2$, PtTe$_2$ and PtSe$_2$, were predicted and found to display tilted Dirac cones \cite{PtTe2, PdTe2, PhysRevB.96.041201, PhysRevB} relative to their Fermi level. For these deemed Dirac type-II materials, the Hamiltonian consists of the type-I linear Hamiltonian plus an additional, momentum dependent, term that breaks Lorentz invariance and leads to quasiparticles displaying a momentum-energy relation that depends on the direction of travel.

In these TMDs the existence of symmetry protected Dirac type-II points in their electronic dispersion, was confirmed through angle-resolved photoemission spectroscopy (ARPES) \cite{PhysRevLett.120.156401,article,article1,article2,PhysRevMaterials.1.074202} and also quantum oscillatory measurements \cite{PhysRevB.97.235154,PhysRevB.96.041201,Yang_2018,PhysRevB.97.245109}. However, not only their Dirac nodes are located relatively far from the Fermi level but the associated Dirac like quasiparticles coexist with topologically trivial carriers which hinders their potential for applications. Still, these compounds offer the possibility of displacing the Fermi level towards the valence bands, where the nodes are located, via chemical doping \cite{Fu_2019,Fei_2018}. However, this approach introduces disorder and may even affect the structural stability of these systems.

In contrast, NiTe$_2$ was predicted to host a Dirac type-II point but very close to the Fermi level\cite{Zhang_2020,doi:10.1021/acs.chemmater.8b02132}. Recent publications claim to find experimental evidence for its existence via the mapping of its electronic band structure through ARPES measurements \cite{PhysRevB.100.195134} or extracting its Berry phase from magnetization measurements \cite{doi:10.1021/acs.chemmater.8b02132}. However, detailed information about the Fermi surface topography, extracted from the angular dependence of the quantum oscillations, along with a comparison with density functional theory (DFT) calculations is still lacking.

In NiTe$_2$, superconductivity was found through the application of hydrostatic pressure displaying a maximum transition temperature at $T_c = 7.5$ K \cite{li2019pressureinduced}, which could be unconventional in character given the topological nature of NiTe$_2$. And although this compound is not superconducting in the bulk, its monolayer was predicted to become superconducting with a $T_c \sim 5.7$ K that might increase up to $\sim 11$ K upon Li intercalation in bilayers \cite{NiTe2SC}. Notice that two-Dimensional NiTe$_2$, i.e. down to a monolayer, can be synthesized via chemical vapor deposition \cite{doi:10.1021/jacs.8b08124}, which offers the possibility of studying unconventional and perhaps topological superconductivity in the monolayer limit.

Given that the precise knowledge of the electronic structure is crucial for unveiling the topological character of NiTe$_2$, and also for predicting its superconducting pairing symmetry, here we report a detailed study on the topography of its Fermi surface through the angular dependence of the de Haas-van Alphen effect. The geometry of its Fermi surface is in broad agreement with Density functional theory calculations, albeit we also observe some mild discrepancies. Although High quality single-crystals grown through a Te flux method display relatively high transport mobilities, we conclude that it is not possible to extract a topologically non-trivial Berry phase from the detected dHvA orbits.
Similar results, confined to lower fields and a restricted angular range for the field orientation, were published in Ref. \onlinecite{doi:10.1021/acs.chemmater.8b02132}. This limited their ability to compare band structure calculations with their experimental results.
Here, we provide additional information concerning, for example, the value and anisotropy of the Land\'{e} \emph{g}-factor, the correct behavior of the magnetic susceptibility implying  the absence of localized magnetic moments, and re-discuss the Berry phase for a specific orbit on the Fermi surface showing that it displays a topologically trivial value in contrast to the claims in Ref. \onlinecite{doi:10.1021/acs.chemmater.8b02132}.

\begin{figure}[tb]
\begin{center}
\includegraphics{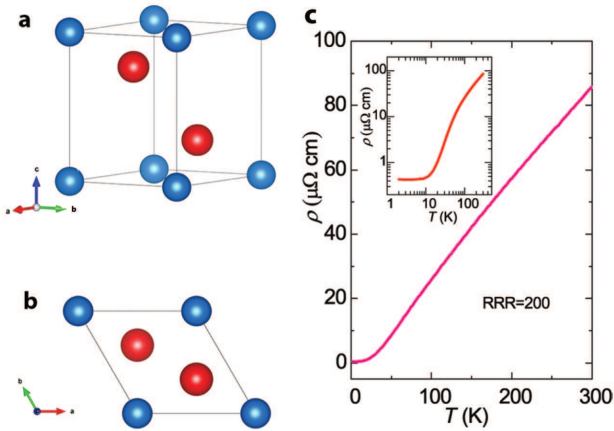}
\caption{\label{fig:epsart} (a) Lateral and (b) top perspective of the unit cell of NiTe$_2$. Ni atoms are depicted through blue spheres while Te atoms are depicted as red spheres. (c) Resistivity $\rho$ as a function of the temperature $T$ for a NiTe$_2$ single crystal. Inset: $\rho(T)$ in a $\log$-$\log$ scale.}
\end{center}
\end{figure}

\begin{figure*}
\begin{center}
\includegraphics[width = 16cm]{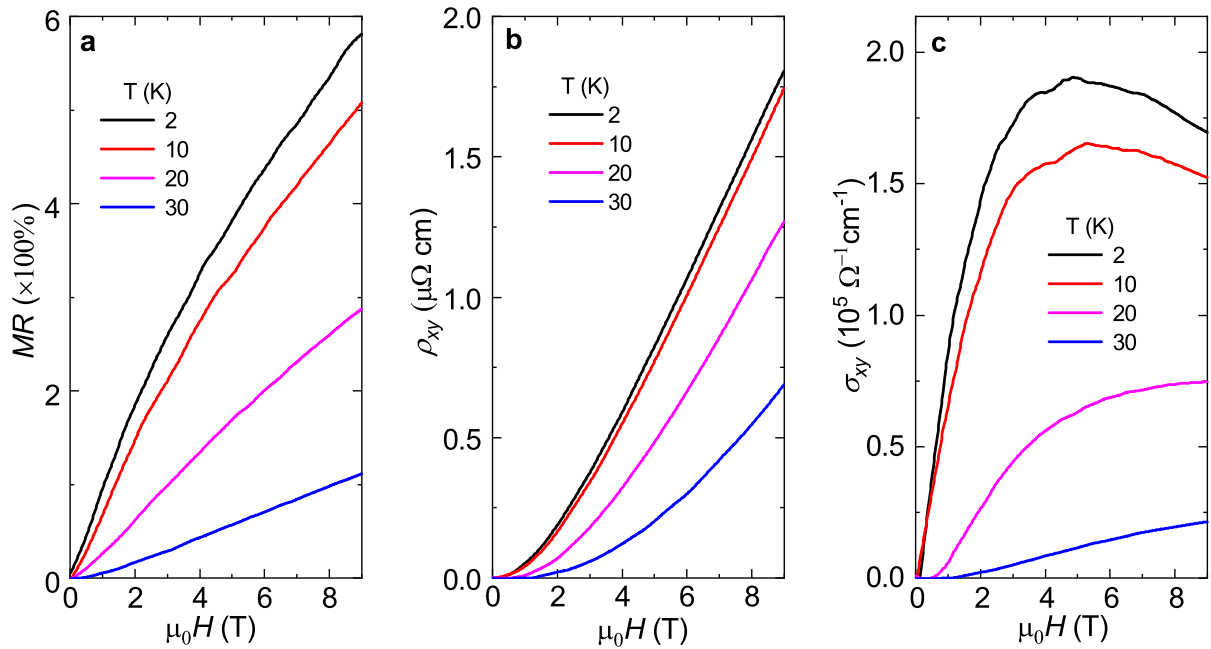}
\caption{The (a) Magnetoresistivity, (b) Hall resistivity and (c) Hall conductivity as functions of the magnetic field $\mu_0H$ applied along the inter-planar direction for a high quality NiTe$_2$ single-crystal at several temperatures. Notice the sub-linear behavior of $\rho(\mu_0H)$ which prevents the extraction of carrier mobilities through the conventional two-band model. However, the maximum in $\sigma_{xy}$ implies an average transport mobility of $\sim 2174$ cm$^2$/Vs in a plane perpendicular to the field.}
\end{center}
\end{figure*}

\begin{figure*}
\begin{center}
\includegraphics{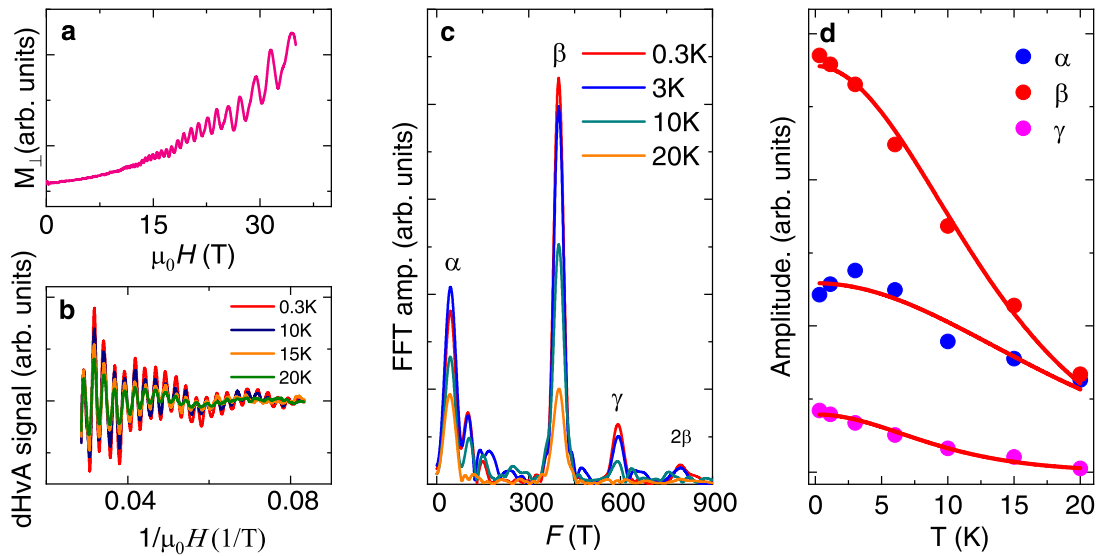}
\caption{\label{fig:wide}(a) Transverse component of the magnetization $M_\perp$ for a NiTe$_2$ single-crystal, as extracted from the magnetic torque $\tau$, as a function of the magnetic field $\mu_0H$. This trace was collected under an external field applied nearly along the \emph{c}-axis at a temperature $T = 0.3$ K. (b) Superimposed oscillatory signal, or the dHvA-effect obtained after subtracting the smoothly varying background, for several temperatures between $T = 0.3 $ K and 20 K. (c) Corresponding fast Fourier transforms (FFTs) of the oscillatory signals. Peaks, corresponding to extremal cross-sectional areas of the Fermi surface, are labeled with Greek letters. (d) Amplitude of the main peaks observed in the FFT spectra as a function of $T$, where solid lines represent fits to the temperature damping factor in the Lifshitz-Kosevich formalism. For the sake of clarity these amplitudes were multiplied by distinct re-scaling factors.}
\end{center}
\end{figure*}

\section{Experimental results}
Single crystals of NiTe$_2$ were grown via Te flux: Ni powders (99.999\% Alfa Aesar) and Te shots (99.999\% Alfa Aesar) with an atomic ratio of 1:20 were sealed in an evacuated quartz ampule and subsequently heated up to 1000 $^\circ$C and held at that temperature for 4 h. Then the ampule was slowly cooled to 525 $^\circ$C at a rate of 1 $^\circ$C/h and subsequently centrifuged.  The as-harvested single crystals were annealed for a few days under a temperature gradient to remove the residual excess Te. The composition and phase purity of the as grown crystals were confirmed by powder and single-crystal x-ray diffraction, see Table S1 and also Fig. S1 \cite{SI}. Single crystal x-ray diffraction reveals a small amount of interstitial Ni which should act as magnetic impurities. However, magnetic susceptibility measured in a commercial Squid magnetometer and in another crystal indicate that this compound is a Pauli paramagnet, see Fig. S2 \cite{SI}. For this particular crystal, the absence of a Curie-Weiss tail at low temperatures suggests that not all crystals contain a sizeable fraction of interstitial Ni. This observation contrasts with the conclusions in Ref. \onlinecite{doi:10.1021/acs.chemmater.8b02132} which claims that NiTe$_2$ would display Curie-Weiss susceptibility and thus break time-reversal symmetry. Conventional magnetotransport experiments were performed in a physical property measurement system (PPMS) using a standard four-terminal method under magnetic fields up to $\mu_0H$ = 9 T and temperatures as low as $T$ = 2 K.  Measurements of the dHvA effect were performed in both a resistive Bitter magnet under continuous fields up to $\mu_0 H$ = 31 T and temperatures as low as $T\simeq$ 0.3 K, and a superconducting magnet coupled to a dilution fridge providing fields up to $\mu_0 H$ = 18 T. Measurements of the dHvA effect were performed via a torque magnetometry technique or by capacitively measuring the deflection of a Cu-Be cantilever with the sample attached to it.
Band structure calculations were performed through the Wien2k implementation \cite{SCHWARZ200271} of Density Functional Theory, using the Perdew-Burke-Ernzerhof (PBE) exchange correlation functional \cite{PhysRevLett.77.3865} in combination with a dense $k$-mesh of 22$\times $8$\times $8 $k$-points and a cutoff $RK_{\text{max}}$ of 7.5. The Fermi surfaces were generated using a $k$-point mesh of 51$\times $51$\times $40 and were visualized using the XCrysden software\cite{KOKALJ1999176}. These calculations were replicated through the QuantumExpresso \cite{QE} implementation of the DFT, finding minor shifts in the position of the bands i.e. in the order of $\sim 20$ meV.
The angular dependence of the extremal cross-sectional areas of the Fermi surface, which are related to the observed de Haas-van Alphen frequencies through the Onsager relation,
was calculated using the SKEAF code\cite{ROURKE2012324}.
\begin{figure}[b]
\begin{center}
\includegraphics[width = 8.6 cm]{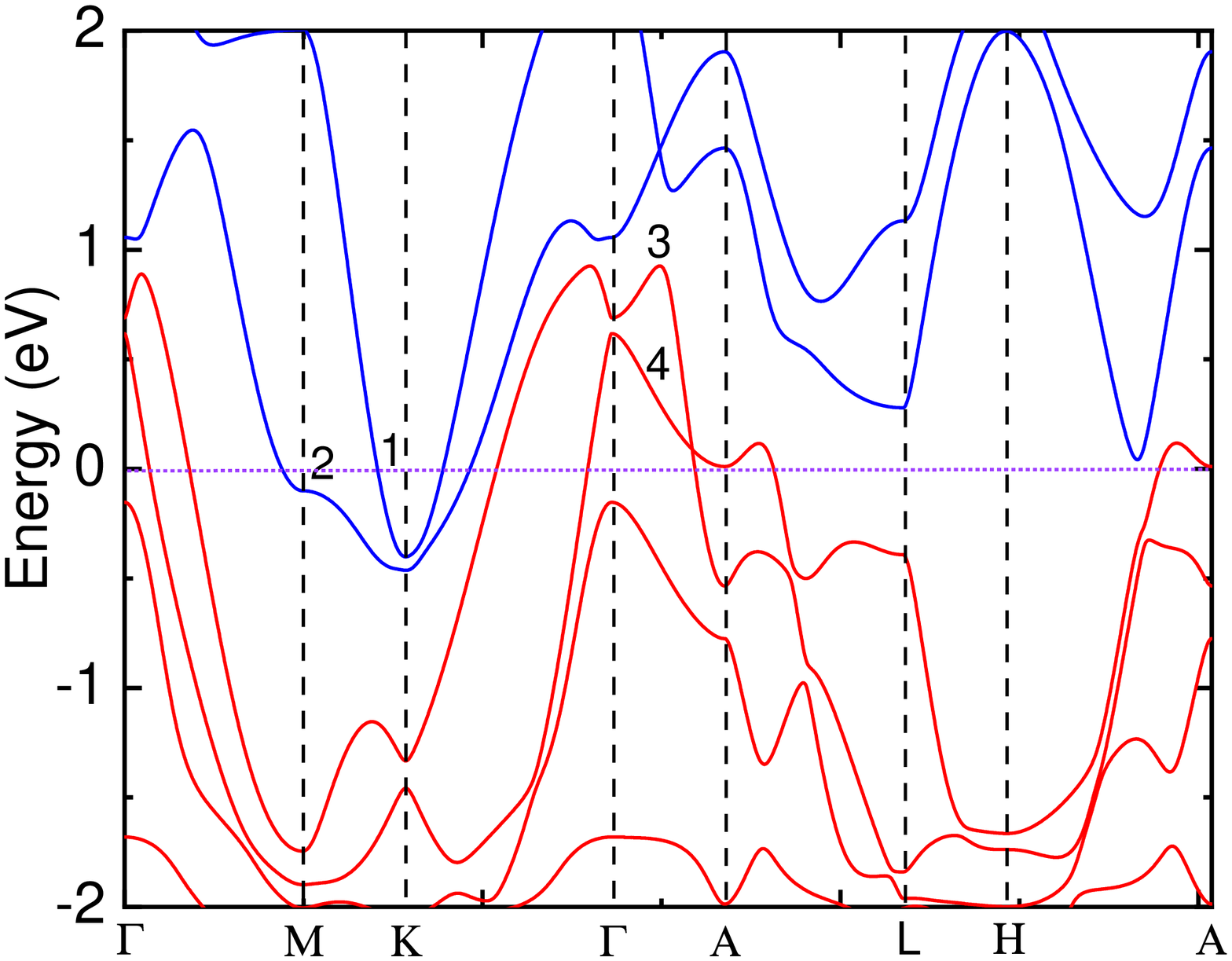}
\caption{\label{fig:epsart} Electronic band structure of NiTe$_2$ including spin-orbit coupling (SOC). A transversal cut of the Dirac type-II cone is observed along the $\Gamma$ to A direction with the node located at $\sim$ 85 meV above the Fermi level. This node is located slightly above 100 meV in the QuantumExpresso \cite{QE} implementation of DFT.
Numbers indicate those bands crossing $\varepsilon_F$ (indicate by purple line).}
\end{center}
\end{figure}

\begin{figure*}
\begin{center}
\includegraphics{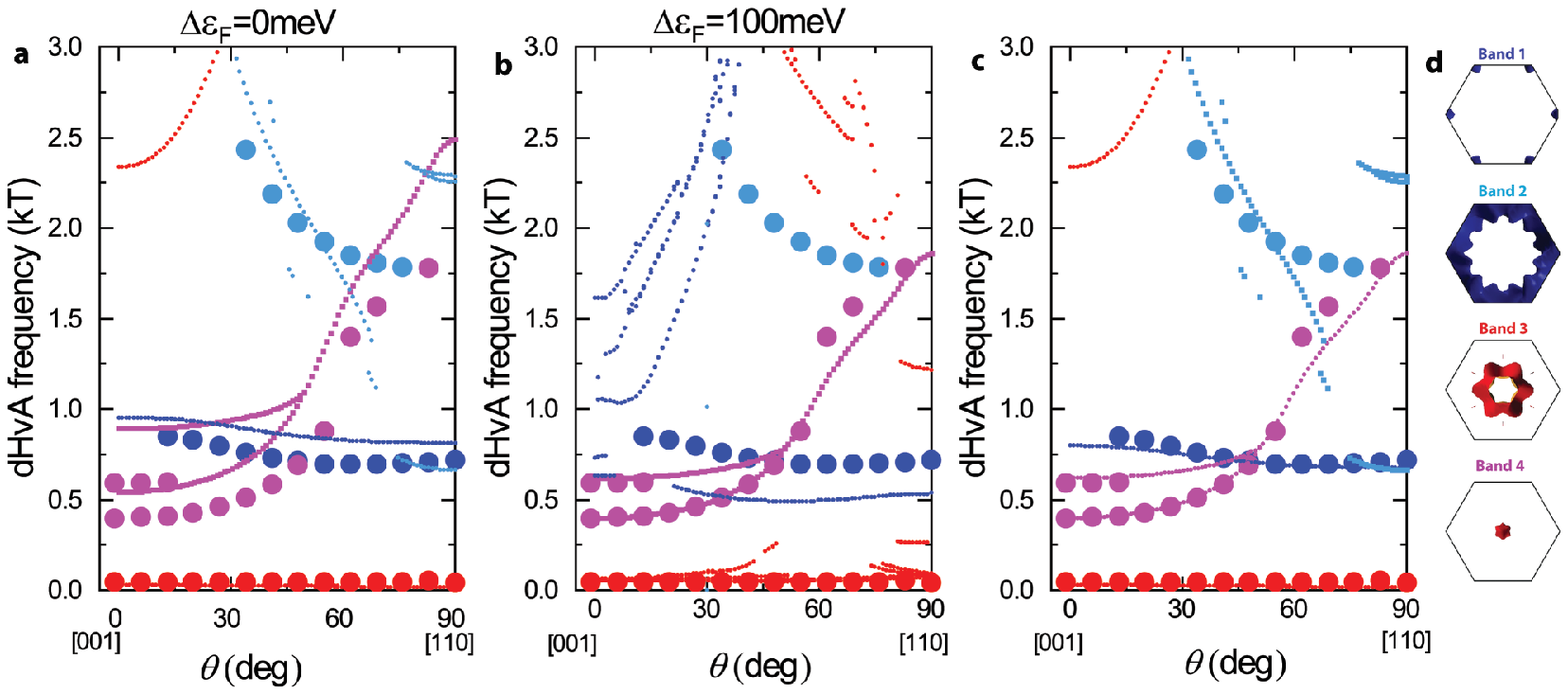}
\caption{\label{fig:wide}Quantum oscillatory frequencies $F$ as a function of the angle $\theta$ between $\mu_0H$ and the \emph{c}-axis of the crystal for different placements of the Fermi level. Larger markers depict the experimental frequencies while small dots depict the angular dependence of $F(\theta)$ according to the DFT calculations. Bands 1 (dark blue) and 2 (light blue) yield electron-like Fermi surface sheets, while bands 3 (red) and 4 (magenta) lead to hole pockets. (a) $F(\theta)$ for the position of the Fermi energy $\varepsilon_F$ resulting from the DFT calculations. (b) $F(\theta)$ with $\varepsilon_F$ shifted by 100 meV. (c) $F(\theta)$ obtained by separately shifting bands 1 and 4 with respect to $\varepsilon_F$, which leads to a better agreement between experiments and calculations. (d) The corresponding Fermi surfaces within the hexagonal first Brillouin zone for the bands in panel (c).}
\end{center}
\end{figure*}

Similarly to PdTe$_2$ and PtTe$_2$, NiTe$_2$ crystallizes in a centro-symmetric trigonal crystallographic structure with space-group $P\bar{3}m1$, whose unit cell is depicted in Figs. 1(a) and 1(b). This structure is inversion symmetric hence the bands are Kramers degenerate. Resistivity as a function of the temperature for a NiTe$_2$ single crystal in the absence of an external magnetic field is shown in the Fig. 1 (c). $\rho(T)$ displays metallic behavior over the entire temperature range. The sample also shows a relatively large residual resistivity ratio (RRR) $\rho$($T$ = 300K)/$\rho$($T$ = 2K) $\simeq$ 200 and a small residual resistivity $\rho \simeq 0.4\ \mu\Omega$cm, which are comparable to values reported earlier \cite{doi:10.1021/acs.chemmater.8b02132}. Although this RRR value is not particularly high, when compared to other layered transition metal dichalcogenides like MoTe$_2$\cite{PhysRevB.94.121101} or WTe$_2$\cite{1}, the resulting residual resistivities are slightly lower, which combined with the presence of quantum oscillations, confirms that these crystals are of high quality.

Under a transverse field ($\mu_0H\parallel$ \emph{c}), the crystal shows sub-linear magnetoresistivity as seen in Fig. 2 (a), while linear magnetoresistance is observed in samples displaying low RRR (see, Figs. 2(a) and 2(b)). A large, linear, and positive magnetoresistivity was already reported decades ago for Ag based chalcogenides \cite{Ag_chalcogenides} and subsequently explained by Abrikosov \cite{Abrikosov} in terms of gapless semiconductors characterized by linearly dispersing quasiparticles analogous to those in Dirac systems. More recently, linear magnetoresistivity has been observed in the parent compounds of the Fe based superconductors \cite{Ba(FeAs)2,Terashima}, which are predicted to display Dirac nodes in their electronic dispersion, and in Dirac semimetallic systems such as Cd$_3$As$_2$ \cite{Cd3As2}.
The magnetoresistivity from Cd$_3$As$_2$ single crystals that display the highest mobilities do display deviations with respect to linearity \cite{Cd3As2}. However, one does not observe sub-linear behavior in field as seen here for the best NiTe$_2$ single-crystals. To explain this effect, it would require a precise understanding on the scattering mechanisms dominating carrier transport in NiTe$_2$.

Unlike the anomalously large magnetoresistivity reported for other topological semimetals, for NiTe$_2$ it increases by merely 600 \% at $T$ = 2 K under $\mu_0H$ = 9 T, probably due to the comparatively large carrier density in this material. For samples displaying a lower RRR we observe no saturation in $\rho (\theta, \mu_0 H)$ under fields all the way up to 31 T (see, Fig. S3(b) \cite{SI}).
\begin{figure*}
\begin{center}
\includegraphics[width = 14 cm]{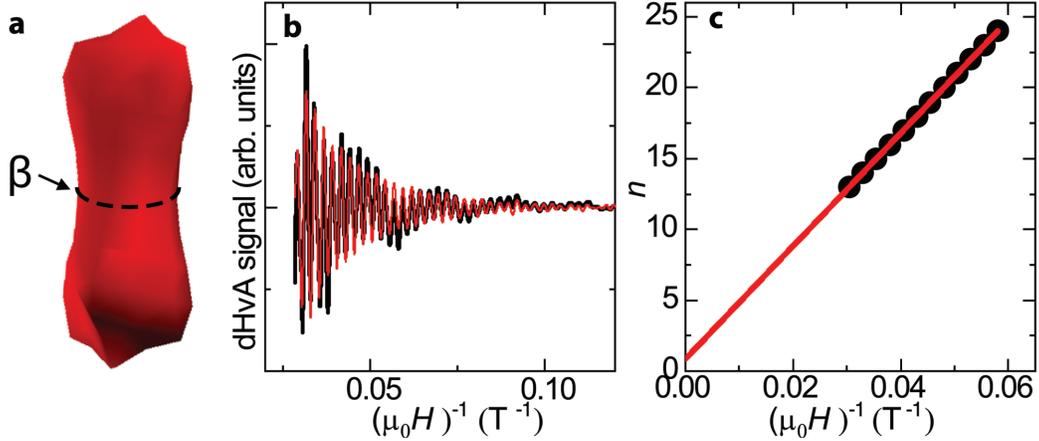}
\caption{\label{fig:wide}(a) Lateral view of the Fermi surface resulting from band 4. (b) Black line represents the raw oscillatory dHvA-signal for NiTe$_2$ at $\theta=0^\circ$. The frequency related to the $\beta$ orbit (neck or minimal cross-sectional area) can be separated by applying a band-pass filter. Solid red line corresponds to the oscillatory component associated solely to the $\beta$ orbit after the application of the band-pass filter. (c) Landau fan diagram for the $\beta$ orbit.}
\end{center}
\end{figure*}
To further evaluate the electrical transport properties of NiTe$_2$, we conducted Hall-effect measurements under fields up to $\mu_0 H$ = 9 T and temperatures between 2 and 300 K. The results are displayed in Fig. 2(b). The Hall resistivity shows nonlinear behavior at low temperatures, especially at low fields, which is a clear indication for multi-band transport with contributions from the various electron and hole-like sheets of the Fermi surface. Given the linear or the sub-linear dependence of the magnetoresistivity with respect to magnetic field, it is not possible to estimate the carrier concentrations $n_{e,h}$ and their mobilities $\mu_{e,h}$ by simultaneously fitting the magnetoresistivy and the Hall-effect to a simple two-band model. However, this approach was used by Ref. \onlinecite{doi:10.1021/acs.chemmater.8b02132} that fits only the Hall data to extract carrier mobilities and concentrations, when in reality conventional transport theory is clearly not applicable to this compound.
Instead, we estimated the mean transport mobility in a plane perpendicular to the external field from the maximum in the Hall conductivity shown in Fig. 2(c), obtaining $\mu_m\simeq$ 2174 cm$^2$V/s at $T$ = 2 K. This value is considerably higher than the mobilities reported in Ref. \cite{doi:10.1021/acs.chemmater.8b02132} from the aforementioned fittings. The observed linear or sub-linear magnetoresistivity when combined with the relatively high mobilities, suggests that Dirac-like quasiparticles  dominate the transport properties of NiTe$_2$, despite their coexistence with topologically trivial carriers.

The magnetic torque $\tau=V \mu_0H \times M$, where $V$ represents the volume of the sample, is proportional to the component of the magnetization perpendicular to the external field ($M_\perp$). Figure 3(a) displays a typical $M_\perp$ trace as a function of magnetic field, collected at $T = 0.3$ K when $\theta \sim 1^\circ$, where $\theta$ = $0^\circ$ and $90^\circ$ correspond to fields along the \emph{c}-axis and the \emph{ab}-plane, respectively. The pure oscillatory or de Haas - van Alphen signal, obtained after subtracting the torque background is shown in Fig. 3(b). The fast Fourier transforms (FFT) of the dHvA signal superimposed onto $M_\perp$ are shown in Fig. 3(c). Several peaks are observable in the FFT spectra at frequencies $F$ between 0 and 900 T, each associated to a Fermi surface cross-sectional area $A$ according to the Onsager relation, $$F = \frac{\hbar}{2\pi e}A, \qquad(1)$$ where $\hbar$ is the reduced Planck constant and $e$ the electron charge. In Fig. 3(d), we plot the amplitude of the main peaks observed in Fig. 3(c) as a function of the temperature, where red lines are fits to the Lifshitz-Kosevich (LK)\cite{shoenberg_1984} temperature damping factor, or $\lambda T/ \sinh (\lambda T)$ where $\lambda = 2\pi^2k_B\mu T /\hbar e B$ and $\mu$ is the carrier effective mass in units of the free electron mass, from which we extract the effective masses associated to each frequency.  We also measured the $M_\perp$ for NiTe$_2$ as a function of $\mu_0H$  for several $T$s when $\theta = 55^\circ$, see all panels in Fig. S4 \cite{SI}. The extracted effective masses for both field orientations are summarized in Table I. We extracted very light effective masses for both field orientations, i.e. ranging between 0.094 $m_0$ and 0.303 $m_0$ for those orbits characterized by frequencies inferior to 900 T. A higher frequency $F_{\varepsilon}$= 1962 T  yielded an effective mass $\mu_{\varepsilon}$= 0.610 $m_0$.

Figure 4 shows the electronic band structure of NiTe$_2$ with the inclusion of spin orbital coupling. Four bands cross the Fermi level, the two electron-like bands are displayed in blue color, while the hole bands are plotted in red. Both hole bands intersect along the $\Gamma$ - A direction at energies slightly above $\varepsilon_F$, thus producing the Dirac type-II cone, which is located at the reciprocal lattice vectors $k^{\text D} = (0,0, \pm 0.35)$ \cite{doi:10.1021/acs.chemmater.8b02132, PhysRevB.100.195134} within the FBZ.

The goal of the this study is to compare the topography of the Fermi surface measured experimentally through the angular dependence of the dHvA-effect, with those predicted by the DFT calculations. A good agreement between both would validate these calculations, and hence support the existence of a tilted Dirac cone in NiTe$_2$.
Therefore, the FFT spectra of the oscillatory dHvA signal collected at $T=2$ K and for various angles is plotted in Fig. S5 \cite{SI}. Oscillations in the torque data are observable over the entire angular range.  Notice that it reveals two spin-zeros for the $\beta$-orbit, or angles where the amplitude of the oscillations approaches zero due to the spin-dephasing factor $R_s = \cos(\pi g \mu /2m_0)$ in the Lifshitz-Kosevich formula \cite{shoenberg_1984}. The cosinusoidal term reaches zero value whenever $\pi g \mu/2m_0 = (2n+1)\pi/2$. Here, $\mu$ is the effective mass in units of the free electron mass and $g$ the Land\'{e} $g$-factor. If one assumed $n = 0$ for the first spin-zero and $n=1$ for the second, one would obtain a large Land\'{e} $g$-factor of 4.74 for $\theta \sim 23^{\circ}$ and $g = 14.22$ for $\theta \sim 53^{\circ}$ implying a pronounced anisotropy by a factor of 3. The origin of such large values for the $g$-factor remains to be understood, although we have observed similar values for the Dirac type-II candidates \emph{M}Al$_3$ \cite{Chen} (where, $M =$ V, Nb or Ta) suggesting perhaps a role for electronic correlations in these systems.

\begingroup
\squeezetable
\renewcommand{\baselinestretch}{1.5}
\begin{table}
\caption {\label{tab:table1} Effective masses of NiTe$_2$ from magnetization measurements, orbits or frequencies indicated with an $\ast$ were collected at an angle $\theta = 55^\circ$. Masses with a ``$B$" subscript indicate calculated band masses.}
\begin{ruledtabular}
\begin{tabular}{ccccccccc}
 \multicolumn{4}{c}{$\theta=0^\circ$}  & \multicolumn{4}{c}{$\theta=55^\circ$}    \\
\cline{1-4} \cline{5-9}
Orbit    &  $F$(T)   & $\mu/m_0$ & $\mu_B/m_0$
	&  Orbit    &  $F$(T)  & $\mu/m_0$ & $\mu_B/m_0$  &Band                                              \\    \hline
 $\alpha$   &  44   & 0.146 & 0.096&$\alpha^*$      & 47   & 0.094&0.075  &3\\
 $\beta$    &  398  & 0.210 & 0.154&$\beta^*$       & 802  & 0.212& 0.437 &4\\
 $\gamma$   &  591  & 0.303 & 0.300      &          &      &       & &4\\
            &       &       & &$\delta^*$      & 706  & 0.202&0.265  &2\\
            &       &       & &$\varepsilon^*$ & 1962 & 0.610 & 1.121&1\\
\end{tabular}
\end{ruledtabular}
\end{table}
\endgroup

Now we proceed to our objective of comparing the angular dependence of the dHvA frequencies with that of the cross-sectional areas of the Fermi surface predicted by DFT, as illustrated by Fig. 5. The experimental frequencies are plotted through larger markers whereas the theoretically predicted ones are indicated by small markers. Figure 5(a) displays the DFT frequencies for the position of $\varepsilon_F$ resulting from the DFT calculations. We observe mild as well as somewhat pronounced band dependent disagreements between calculations and experiments. However, if we followed the ARPES approach in Ref. \onlinecite{PhysRevB.100.195134} and shifted $\varepsilon_F$ by +100 meV one would achieve a perfect agreement between the DFT calculations and the experimental frequencies for band 4, see Fig. 5(b). However, as clearly  seen such displacement intensifies the disagreement for the other bands.
Therefore, we choose to displace individual bands with respect to $\varepsilon_F$ to improve the agreement. This ad-hoc approach assumes that the exact position of the bands
usually depends on the precise details of the DFT implementation, with these small displacements correcting for the inherent error of DFT.
Moving band 1 downward by 60 meV with respect to $\varepsilon_F$ (as shown in Fig. 5(c)) leads to a better match. Nevertheless, displacing all the bands
by this same amount decreases the overall agreement. Therefore, in Fig. 5(d) the calculated bands 1 and 4 were displaced by -60 meV and 100 meV relative to $\varepsilon_F$ respectively, leading to a very good agreement for the observed FS sheets. However, this independent band shift would lead to the suppression of the Dirac type-II nodes
and hence would be inconsistent with the general good agreement between the band structure collected by ARPES \cite{PhysRevB.100.195134} and the calculations. A similar situation was observed for the Weyl type-II semi metallic candidate MoTe$_2$ where the electronic bands had to be independently displaced to match both the ARPES and the dHvA data \cite{Daniel}. Subsequent theoretical work \cite{PhysRevB.99.035123}, supported by an ARPES study \cite{xu}, pointed to the relevant role played by electronic correlations which
tend to renormalize the dispersion of the electronic bands. Perhaps, the deviations seen here might indicate that electronic correlations are relevant also to NiTe$_2$ although these do not renormalize the carrier effective masses with respect to the band masses according to Table \ref{tab:table1}. Instead, correlations would seem to favor a renormalized Land\'{e} \emph{g}-factor. Finally, Fig. 5(d) shows the corresponding Fermi surfaces resulting from the displaced bands. As shown in Fig 3(c), for magnetic fields oriented along the \emph{c}-axis, there are three distinct frequencies: $\alpha$, $\beta$ and $\gamma$. Based on their angular dependence, the $\alpha$ frequency can be assigned to the very small hole pockets resulting from band 3, while the $\beta$ and $\gamma$ orbits would correspond to the minimum and maximum cross sectional areas of the hole pocket associated with band 4 (see, Fig. 6 (a)).

Given the possible presence of Dirac type-II points in NiTe$_2$, one might expect to detect charge carriers characterized by topologically nontrivial Berry phases. The LK formula\cite{shoenberg_1984} describes the quantum oscillatory phenomena observed in the magnetization $M$ through:
$$\triangle M \propto -B^{1/2}R_TR_DR_S \sin {2\pi \left[\frac{F}{B}-\left(\frac{1}{2}-\phi\right)\right]},\quad (2)$$
where $R_T$ is the previously mentioned temperature damping factor, $R_D$=exp$(-\lambda T_D)$ is the Dingle damping factor, and $R_s$ is the spin dephasing factor. The phase factor $\phi=\phi_B/2\pi-\delta$, contains the Berry phase $\phi_B$ and an additional term that takes values of 0 or $\pm 1/8$ (the sign depends on the cross-sectional area, maxima or minima) for Fermi surfaces of two- or three-dimensional character, respectively. In order to extract the correct phase of the dHvA oscillations, one can use the magnetic susceptibility $\triangle\chi=d(\triangle M)/d(\mu_0 H):$
$$ \triangle\chi\sim{\rm sign}R_s \cos{2\pi\left[\frac{F}{B}- \left(\frac{1}{2}-\phi \right)\right]}, \quad (3)  $$
The Dirac node  located along the $\Gamma$ - A direction, would touch the extreme tip or the extreme bottom of the hole Fermi surface resulting from band 4 depicted Fig. 6(a), if $\varepsilon_F$ was precisely located at the energy of the Dirac type-II point.

As is shown in Fig. 6(a), the $\beta$ frequency corresponds to an orbit around the minimum cross-sectional area of the hole pocket centered around the $\Gamma$-point. It displays the dominant peak in the FFT spectra seen in Fig. 3(c) for $\theta$ $<$ $25^\circ$. It is also well-separated from the other frequencies making it a good
candidate for the evaluation of the Berry-phase of NiTe$_2$, as previously done in Ref. \onlinecite{doi:10.1021/acs.chemmater.8b02132}. For example, one can apply a band pass filter to extract only the oscillations associated with the $\beta$ orbit ($F_\beta$ = 398 T). The result of this procedure is shown in Fig. 6(b). The valleys in the oscillatory signal can be assigned to Landau indices \emph{n}, which produces the Landau fan diagram shown in Fig. 6(c). The phase $\phi$ can be extracted from the intercept of the extrapolation of $n$ as a function of $(\mu_0H)^{-1}$. For $\theta = 0^\circ$, we obtain $\phi = \phi_B/2\pi + 1/8 = 0.85$, corresponding to a Berry phase of $\phi_B=1.45\pi$. Similar results are obtained for other two angles satisfying $\theta < 25^\circ$. However, for a Dirac system one would expect to extract a Berry phase of $\pi$.
This discrepancy is not at all surprising, given that the $\beta$-orbit does not encircle the Dirac node due its position in the $k_z$ plane relative to the position of the Dirac Type-II node at $k_z^{\text D} = \pm 0.35$. This relative position precludes this orbit from probing the texture of the Berry curvature around the Dirac node, as discussed in Ref. \onlinecite{Chen}. In addition, and as previously mentioned, the spin-dephasing factor $R_S$ changes the sign of the amplitude of the oscillatory term as the sample is rotated with respect to the external field, thus affecting the proper extraction of the phase. There is however another frequency associated to band 4, the $\gamma$-orbit, that corresponds to the maximal cross sectional area on the FS sheet shown in Fig. 6(a) which is closer to the nodes. However, its weak amplitude in the FFT spectra makes it nearly impossible to reliably extract its Berry phase.

\section{Summary}
In summary, we unveiled the Fermi surfaces of NiTe$_2$ through measurements of the de Haas-van Alphen effect and compared it to band structure calculations.
The general trends of the detected Fermi surface sheets are relatively well captured by density functional theory calculations. However, the precise position of the
bands relative to the Fermi level had to be independently displaced to achieve an optimal agreement. The need for this ad-hoc procedure, in combination with
light and nearly isotropic experimental effective masses, indicates that density Functional theory fails to capture the finer details of the electronic band-structure
of NiTe$_2$. The Land\'{e} $g$-factor extracted for this system is observed to be large and anisotropic as previously observed in several topological semimetals,
pointing to either a possible role for correlations, or a large orbital angular momentum for these systems.  The anomalous magnetoresistivity displayed by this system in combination with the relatively high mobilities, suggest that Dirac-like quasiparticles dominate its transport properties, despite their coexistence with topologically trivial carriers. It remains to be understood how these Lorentz invariance breaking quasiparticles might influence the superconductivity observed in this compound under pressure.\\

\section{Acknowledgments}
We acknowledge R. E. Baumbach for the assistance with the magnetic susceptibility measurements. This work was supported by DOE-BES through award DE-SC0002613. W.Z. was partially supported by NSF through NSF-DMR-1807969. The NHMFL is supported by NSF through NSF-DMR-1644779 and the State of Florida.

\end{document}